\documentclass[aps,twocolumn,superscriptaddress,preprintnumbers,showpacs,prl]{revtex4}
\usepackage{graphicx,amssymb,epsfig,amsmath}
\usepackage{dcolumn}% Align table columns on decimal point
\usepackage{bm}% bold math
\usepackage{epsfig}
\begin{document}
\draft \preprint{Beaujour {\it et al}.}

\title {Ferromagnetic resonance linewidth in ultrathin films with perpendicular magnetic anisotropy}
\author{J-M.~Beaujour}
\affiliation{Department of Physics, New York University, 4 Washington Place, New York, New York 10003, USA}
\author{D.~Ravelosona}
\affiliation{Institut d'Electronique Fondamentale, UMR CNRS 8622, Universit$\acute{e}$ Paris Sud, 91405 Orsay Cedex, France}
\author{I.~Tudosa}
\affiliation{Center for Magnetic Recording Research, University of California, San Diego, La Jolla, California 92093-0401, USA}
\author{E.~Fullerton}
\affiliation{Center for Magnetic Recording Research, University of California, San Diego, La Jolla, California 92093-0401, USA}
\author{A.~D.~Kent}
\affiliation{Department of Physics, New York University, 4 Washington Place, New York, New York 10003, USA}

\date{\today}

%#############################################
\begin{abstract}
Transition metal ferromagnetic films with perpendicular magnetic anisotropy (PMA) have ferromagnetic resonance (FMR) linewidths that are one order of magnitude larger than soft magnetic materials, such as pure iron (Fe) and permalloy (NiFe) thin films. A broadband FMR setup has been used to investigate the origin of the enhanced linewidth in Ni$|$Co multilayer films with PMA. The FMR linewidth depends linearly on frequency for perpendicular applied fields and increases significantly when the magnetization is rotated into the film plane. Irradiation of the film with Helium ions decreases the PMA and the distribution of PMA parameters. This leads to a great reduction of the FMR linewidth for in-plane magnetization. These results suggest that fluctuations in PMA lead to a large two magnon scattering  contribution to the linewidth for in-plane magnetization and establish that the Gilbert damping is enhanced in such materials  ($\alpha \approx 0.04$, compared to $\alpha \approx 0.002$ for pure Fe).
\end{abstract}
%#############################################

 \maketitle

%#############################################
%\section{Introduction}
%#############################################
Magnetic materials with perpendicular magnetic anisotropy (PMA) are of great interest in  information storage technology, offering the possibility of smaller magnetic bits \cite{Thomson2006} and more efficient magnetic random access memories based on the spin-transfer effect \cite{Mangin2006}. They typically are multilayers of transition metals (e.g., Co$|$Pt, Co$|$Pd, Ni$|$Co) with strong interface contributions to the magnetic anisotropy \cite{Daalderop1992}, that render them magnetically hard. In contrast to soft magnetic materials which have been widely studied and modeled \cite{Heinrich2005,Scheck2007,Gilmore2007,Brataas2008}, such films are poorly understood. Experiments indicate that there are large distributions in their magnetic characteristics, such as their switching fields \cite{Thomson2006}. An understanding of magnetization relaxation in such materials is of particular importance, since magnetization damping determines the performance of magnetic devices, such as the time-scale for magnetization reversal and the current required for spin-transfer induced switching \cite{Sun2000,Mangin2006}.

Ferromagnetic resonance (FMR) spectroscopy provides information on the magnetic damping through study of the linewidth of the microwave absorption peak, $\Delta H$, when the applied field is swept at a fixed microwave frequency. FMR studies of thin films with PMA show very broad linewidths, several 10's of mT at low frequencies ($\lesssim 10$ GHz) for polycrystalline alloy \cite{Clinton2008}, multilayer \cite{Yuan2003} and even epitaxial thin films \cite{BenYoussef1999}. This is at least one order of magnitude larger than the FMR linewidth found for soft magnetic materials, such as pure iron (Fe) and permalloy (FeNi) thin films \cite{Scheck2007}. Further, it has recently been suggested that the FMR linewidth of perpendicularly magnetized CoCrPt alloys cannot be explained in terms of Landau-Lifshitz equation with Gilbert damping \cite{Mo2008}, the basis for understanding magnetization dynamics in ferromagnets:
\begin{eqnarray}
{{\partial {\bf M}} \over{\partial t} }=-\gamma \mu_0{\bf M} \times {\bf H_\text{eff}} +{\alpha \over M_s} {\bf M}
\times {{\partial {\bf M}} \over{\partial t}} \:.
\label{llg}
\end{eqnarray}
Here ${\bf M}$ is the magnetization and $\gamma$$=|g\mu_B/\hbar|$ is the gyromagnetic ratio. The second term on the right is the damping term, where $\alpha$ is the Gilbert damping constant. This equation describes precessional motion of the magnetization about an effective field ${\bf H_\text{eff}}$, that includes the applied and internal (anisotropy) magnetic fields, which is damped out at a rate determined by $\alpha$. The absorption linewidth (FWHM) in a fixed frequency field-swept FMR experiment is given by $\mu_0\Delta H = 4\pi \alpha f/\gamma$, i.e., the linewidth is proportional to the frequency with a slope determined by $\alpha$. This is the homogeneous or intrinsic contribution to the FMR linewidth. However, experiments show an additional frequency independent contribution to the linewidth: 
\begin{equation}
\label{Eq:linewidth} \Delta H = \Delta H_0 + \frac{4 \pi \alpha}{\mu_0\gamma} f,
\end{equation}
where $\Delta H_0$ is referred as the inhomogeneous contribution to the linewidth. 

The inhomogeneous contribution is associated with disorder. First, fluctuations in the materials magnetic properties, such as its anisotropy or magnetization, lead to a linewidth that is frequency independent; in a simple picture, independent parts of the sample come into resonance at different applied magnetic fields. Second, disorder can couple the uniform precessional mode ($k=0$), excited in an FMR experiment, to degenerate finite-$k$  ($k \neq 0$) spin-wave modes. This mechanism of relaxation of the uniform mode is known as two magnon scattering (TMS) \cite{Sparks1964}. TMS requires a spin-wave dispersion with finite-$k$ modes that are degenerate with the $k=0$ mode that only occurs for certain magnetization orientations.  

In this letter we present FMR results on ultra-thin Ni$|$Co multilayer films and investigate the origin of the broad FMR lines in films with PMA. Ni$|$Co multilayers were deposited between Pd$|$Co$|$Pd layers that enhance the PMA and enable large variations in the PMA with Helium ion irradiation \cite{Stanescu2008}. The films are $|$3nm Ta$|$1nm Pd$|$0.3nm Co$|$1nm Pd$|$[0.8nm Ni$|$0.14nm Co]$\times$3$|$1nm Pd$|$0.3nm Co$|$1nm Pd$|$0.2nm Co$|$3nm Ta$|$ deposited on a Si-SiN wafers using dc magnetron sputtering and were irradiated using 20 keV He$^+$ ions at a fluence of 10$^{15}$ ions/cm$^2$. The He$^+$ ions induce interatomic displacements that intermix the Ni$|$Co interfaces leading to a reduction of interface anisotropy and strain in the film. The magnetization was measured at room temperature with a SQUID magnetometer and found to be $M_s \simeq 4.75 \times 10^5$ A/m.

\begin{figure}[t]
\begin{center}\includegraphics[width=7cm]{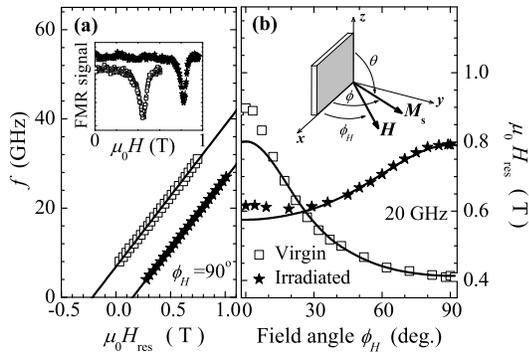}
\vspace{-4 mm}\caption{\label{fig1} a) The frequency dependence of the resonance field with the applied field perpendicular to the film plane. The solid lines are fits using Eq. \ref{Eq:Hres}. Inset: FMR signal of the virgin and irradated films at 21 GHz. b) The resonance field as a function of applied field angle at 20 GHz. The solid lines are fits to the experimental data points. The inset shows the field geometry.}
\vspace{-6 mm}
\end{center}
\end{figure}

FMR studies were conducted from 4 to 40 GHz at room temperature with a coplanar waveguide as a function of the field angle to the film plane. The inset of Fig. \ref{fig1}b shows the field geometry. The parameters indexed with `$\perp$' (perpendicular) and `$\parallel$' (parallel) refer to the applied field direction with respect to the film plane. The absorption signal was recorded by sweeping the magnetic field at constant frequency \cite{Beaujour2007}. FMR measurements were performed on a virgin film (not irradiated) and on an irradiated film. 

%#############################################
%\section{Results-FMR position}
%#############################################

Fig. \ref{fig1}a shows the frequency dependence of the resonance field when the applied field is perpendicular to the film plane. The $x$-intercept enables determination of the PMA and the slope is proportional to the gyromagnetic ratio. We take a magnetic energy density:
\begin{equation}
\begin{split}
\label{Eq:Energy} E&= -\mu_0 {\bf M} \cdot {\bf H} + {1 \over 2} \mu_0 M_{\mathrm{s}}^2 \sin^2 \phi\\
&\quad-(K_1+2K_2) \sin^2 \phi + K_2\sin^4 \phi .
\end{split}
\end{equation}
The first term is the Zeeman energy, the second the magnetostatic energy and the last two terms include the first and second order uniaxial PMA constants, $K_1$ and $K_2$. Taking $\mu_0{\bf H_{\rm eff}}=-\delta E/\delta{\bf{M}}$ in Eq. \ref{llg} the resonance condition is:
\begin{equation}
\label{Eq:Hres} 
f={\gamma \over 2 \pi}\left( \mu_0 H_{\mathrm{res}}^{\perp} - \mu_0 M_{\mathrm{s}} + \frac{2
K_1}{M_{\mathrm{s}}}\right).
\end{equation}
From the $x$-intercepts in Fig.~\ref{fig1}a, $K_1=(1.93\pm0.07)\times10^5$~J/m$^3$ for the virgin film and $(1.05\pm0.02)\times10^5$~J/m$^3$ for the irradiated film; Helium irradiation reduces the magnetic anisotropy by a factor of two. Note that in the irradiated film the $x$-intercept is positive ($\mu_0 M_s>2K_1/M_{\mathrm{s}}$). This implies that the easy magnetization direction is in the film plane. The angular dependence of $H_{\mathrm{res}}$ (Fig. \ref{fig1}b) also illustrates this: the maximum resonance field  shifts from in-plane to out-of-plane on irradiation. The gyromagnetic ratio is not significantly changed $\gamma=1.996\pm0.009 \times 10^{11}~\:~$1$/$(Ts) for the virgin  film and $\gamma=1.973\pm0.004 \times 10^{11}~\:~$1$/$(Ts) for the irradiated film (i.e., $g=2.24\pm0.01$).
The second order anisotropy constant $K_2$ was obtained from the angular dependence of the resonance field, fitting $H_{\mathrm{res}}$ versus $\phi_H$ for magnetization angle $\phi$ between $45^{\mathrm{o}}$ and $90^{\mathrm{o}}$. For the virgin film, $K_2=0.11 \times 10^5$~J/m$^3$. Note that when $K_2$ is set to zero, $\chi^2$ of the fit increases by a factor 30. For the irradiated film, $K_2=0.03 \times 10^5$~J/m$^3$. Hence $K_2$ decreases upon irradiation and remains much smaller than $K_1$. The solid line in Fig. \ref{fig1}b is the resulting fit. When the field approaches the in-plane direction, the measured resonance field is higher than the fit. The shift is of the order of $0.1$~T for the virgin film and $0.025$~T for the irradiated film. It is frequency dependent: increasing with frequency. This shift will be discussed further below.

%*********************************************
%\section{Results-FMR linewidth}
%*********************************************

\begin{figure}[b]
\begin{center}
\includegraphics[width=5.5cm]{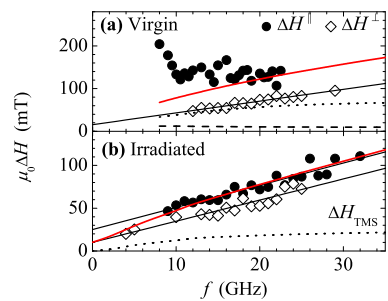}
\vspace{-5 mm}\caption{\label{fig2} The frequency dependence of the FMR linewidth with applied field in-plane and perpendicular to the plane. The solid black lines are linear fits that enable determination of $\alpha$ and $\Delta H_0$ from Eq. \ref{Eq:linewidth}. The dotted lines show the linewidth from TMS and the red lines is the total linewidth.}
\vspace{-6 mm}
\end{center}
\end{figure}
Fig.~\ref{fig2}a shows the frequency dependence of the linewidth (FWHM) for two directions of the applied field. $\Delta H^{\perp}$ of the virgin film increases linearly with frequency consistent with Gilbert damping. Fitting to Eq. \ref{Eq:linewidth}, we find $\alpha=0.044\pm0.003$ and $\mu_0\Delta H_0^{\perp}=15.6\pm3.6$~mT. When the field is applied in the film plane, the linewidth is significantly larger. $\Delta H^{\parallel}$ decreases with increasing frequency for $f \leq10$~GHz and then is practically independent of frequency, at $\approx140\pm20$~mT. However, for the irradiated film, the linewidth varies linearly with frequency both for in-plane and out-of-plane applied fields, with nearly the same slope. 
The Gilbert damping is $\alpha=0.039\pm0.004$. Note that $\mu_0 \Delta H_0^{\parallel}$ is larger than $\mu_0\Delta H_0^{\perp}$ by about $15$~mT.
\begin{figure}
\begin{center}\includegraphics[width=6cm]{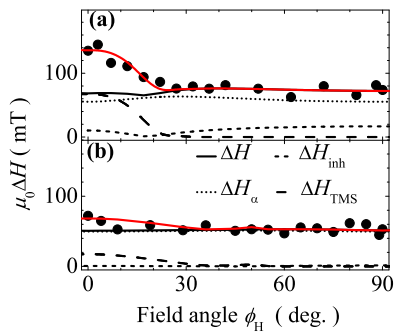}
\vspace{-4 mm}\caption{\label{fig3} Angular dependence of the linewidth at 20 GHz for (a) the virgin and (b) the irradiated film. The solid line ($\Delta H$) is a best fit of the data that includes the Gilbert damping ($\Delta H_{\alpha}$) and the inhomogeneous ($\Delta H_{\mathrm{inh}}$) contributions. Linewidth broadening from TMS ($\Delta H_{\mathrm{TMS}}$) is also shown. The total linewidth is represented by the red line.}
\vspace{-6 mm}
\end{center}
\end{figure}

The angular dependence of the linewidth at $20$~GHz is shown in Fig. \ref{fig3}. The linewidth of the virgin film decreases significantly with increasing field angle up ~30$^{\mathrm{o}}$, and then is nearly constant, independent of field angle. The linewidth of the irradiated film is nearly independent of the field angle, with a relatively small enhancement of $\sim15$~mT close to the in-plane direction. We fit this data assuming that the inhomogeneous broadening of the line is associated mainly with spatial variations of the PMA, specifically local variation in $K_1$, $\Delta H_{\mathrm{inh.}} (\phi_H)= \left|\partial H_{\mathrm{res}}/\partial K_1\right| \Delta K_1$. $\Delta K_1=4\times10^3$~J/m$^3$ for the virgin film and 3$\times 10^2$~J/m$^3$ for the irradiated film, which corresponds to a variation of $K_1$ of 2\% and 0.3\% respectively. Including variations in $K_2$ and anisotropy field direction  do not significantly improve the quality of the fit. Such variations in $K_1$ produce a zero frequency linewidth in the perpendicular field direction, $\mu_0\Delta H_0^{\perp}=16.8$~mT, in excellent agreement with linear fits to the data in Fig. \ref{fig2}.
However, the combination of inhomogeneous broadening and Gilbert damping  {\em cannot} explain the enhanced FMR linewidth observed for in-plane applied fields.
%#############################################
%section{Interpretation}
%#############################################

\begin{figure}[t]
\begin{center}\includegraphics[width=6cm]{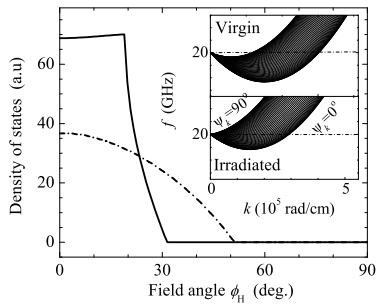}
\vspace{-7 mm}\caption{\label{fig4} The density of spin-waves states degenerate with the uniform mode as a function of field angle at 20 GHz for the virgin film (solid line) and the irradiated film (dashed-dotted line). Inset: Spin wave dispersion when the dc field is in the film plane.}
\label{Fig4}
\vspace{-6 mm}
\end{center}
\end{figure}

The enhanced linewidth observed with in-plane applied fields is consistent with a significant TMS contribution to the relaxation of the uniform mode--the linewidth is enhanced only when finite-$k$ modes equi-energy with the uniform mode are present. We derive the spin-wave dispersion for these films following the approach of \cite{Landeros2008}:
\begin{equation}
\begin{split}
\label{Eq:dispersion} \omega_k ^2&= \omega_{0}^2-{1 \over 2}\gamma ^2 \mu_0 M_s kt (B_{x0} (\cos{2\phi} \\
&\quad+\sin^2{\phi} \sin^2{\psi_k})-B_{y0}\mathrm{sin}^2 \psi_k)+ \gamma^2 D k^2 (B_{x0}+B_{y0}) ,
\end{split}
\end{equation}
where:
\begin{equation}
\begin{split}
\label{Eq:dispersion1}
B_{x0}=&\mu_0H\cos(\phi_H-\phi)- \mu_0 M_{\mathrm{eff}}\mathrm{sin}^2 \phi\\
B_{y0}=&\quad\mu_0 H\cos(\phi_H-\phi)
+ \mu_0 M_{\mathrm{eff}}\cos 2\phi\\
 +\frac{2K_2}{M_s}\sin^2 2\phi.
\end{split}
\end{equation}
The effective demagnetization field is $\mu_0 M_{\mathrm{eff}}=(\mu_0 M_s-\frac{2 K_1}{M_s}-\frac{4K_2}{M_s}\cos^2\phi)$. 
$\omega_0=\gamma \sqrt{B_{x0} B_{y0}}$ is the resonance frequency of the uniform mode. 
$D$ is the exchange stiffness and $t$ is the film thickness. $\psi_k$ is the direction of propagation of the spin-wave in the film plane relative to the in-plane projection of the magnetization.  The inset of Fig. \ref{fig4} shows the dispersion relation for the virgin and the irradiated film for an in-plane applied field at $20$~GHz. For the virgin film, with the easy axis normal to the film plane ($M_{\mathrm{eff}}^{\parallel}<0$) there are degenerate modes available in all directions in $k$-space. For the irradiated film ($M_{\mathrm{eff}}^{\parallel}>0$) degenerate modes are only available when $\psi_k \lesssim 74^{\mathrm{o}}$.

The spin waves density of states, determined from Eq. \ref{Eq:dispersion}, is shown as a function of field angle in Fig. \ref{Fig4} at 20 GHz. The DOS of the virgin film is two times larger than that of the irradiated film at $\phi_H=0$. Note that for both films, the DOS vanishes at a critical field angle that corresponds to a magnetization angle $\phi=45^{\mathrm{o}}$. For the virgin film, the enhancement of $\Delta H$ occurs at $\phi_H\simeq 30^{\mathrm{o}}$ (Fig. \ref{fig3}a), at the critical angle seen in Fig. \ref{Fig4}.

The TMS linewidth depends on the density of states and the disorder, which couples the modes:
\begin{equation}
\label{Eq:TwoMagnon} 
\Delta H_{\mathrm{TMS}} = \left(\frac{\partial H_{\mathrm{res}}}{\partial \omega} \right) \frac{|A_{0}|^2}{2 \pi} \int{C_k(\xi)\delta (\omega_k - \omega_0)} d{\bf k},
\end{equation}
where $A_0$ is a scattering amplitude. $C_k(\xi)=2 \pi \xi^2 /(1+(k \xi)^2)^{3/2}$ is a correlation function, where $\xi$ is correlation length, the typical length scale of disorder. Eq. \ref{Eq:TwoMagnon} is valid in the limit of weak disorder.

We assume that the disorder of our films is associated with spatial variations of the PMA,  $K_1$. Then the magnetic energy density varies as $\Delta E(\vec{r})=-k_1(\vec{r}) M_y^2/M_{\mathrm{s}}^2$, and the scattering probability is \cite{McMichael2004}:
\begin{equation}
\begin{split}
|A_0|^2=&\frac{\gamma^4}{4 \omega_0 ^2} (B_{x0}^2 \sin^4 \phi + B_{y0}^2 \cos^2 2\phi \\
&\quad- 2 (\omega_0/\gamma)^2\sin^2\phi \cos 2\phi) \left({\frac{2 \Delta k_1}{M_s}}\right)^2.
\label{Eq:ScatteringProbability}
\end{split}
\end{equation}
$\Delta k_1$ is the rms amplitude of the distribution of PMA, $k_1(r)$.
Therefore the TMS linewidth broadening scales as the square of  $\Delta k_1$. Since the variations in PMA of the virgin film are larger than that of the irradiated film the linewidth broadening from the TMS mechanism is expected to be much larger in the virgin film, qualitatively consistent with the data.

A best fit of the linewidth data to the TMS model is shown in Fig. \ref{fig3}a. For the virgin film, we find $\xi \approx 44$ nm, approximately four times the film grain size, and $\Delta k_1=9\times 10^{3}$~J$/$m$^3$. The exchange stiffness, $D=2 A/ \mu_0 M_s$ with the exchange constant $A=0.83 \times 10^{-11}$ J/m, is used in the fittings. The cut-off field angle for the enhancement of the field linewidth agrees well with the data (Fig. \ref{fig3}a). For the irradiated film, a similar analysis gives: $\xi=80 \pm 40$ nm and $\Delta k_1=(4\pm2) \times 10^3$~J$/$m$^3$.

TMS is also expected to shift the resonance position \cite{McMichael2004}. For applied fields in-plane and $f=20$~GHz we estimate the resonance field shift to be $\approx 33$ mT. This is smaller than what is observed experimentally ($\approx$ 93 mT). 
The deviations of the fits in Fig.~\ref{fig1}b may be associated with an anisotropy in the gyromagnetic ratio, i.e. a $g$ that is smaller for ${\bf M}$ in the film plane.  Note that if we assume that the $g$-factor is slightly anisotropic ($\sim1$\%), we can fit the full angular dependence of the resonance field of the irradiated film.

We note that the TMS model cannot explain the enhanced linewidth for small in-plane applied fields for the virgin film (Fig. \ref{fig2}a).  The FMR linewidth increases dramatically when the frequency and resonance field decreases. When the applied in-plane field is less than the effective demagnetization field ($-\mu_0 M_{\mathrm{eff}}^{||}=0.31$~T) the magnetization reorients out of the film plane. For frequencies less than about $8$ GHz this leads to two resonant absorption peaks, one with the magnetization having an out-of-plane component for $H_{\mathrm{res}}<-M_{\mathrm{eff}}^{||}$ and one with the magnetization in-plane for $H_{\mathrm{res}}> -M_{\mathrm{eff}}^{||}$. It may be that these resonances overlap leading to the enhanced FMR linewidth.

%section{Summary}
In sum, these results show that the FMR linewidth in Ni$|$Co multilayer films is large due to disorder and TMS as well as enhanced Gilbert damping. The latter is an intrinsic relaxation mechanism, associated with magnon-electron scattering and spin-relaxation due to spin-orbit scattering. As these materials contain heavy elements such as Pd and short electron lifetimes at the Fermi level, large intrinsic damping rates are not unexpected. The results indicate that the FMR linewidth of Ni$|$Co multilayers can be reduced through light ion-irradiation and further demonstrate that the Gilbert damping rate is largely unaffected by irradiation. These results, including the reduction of the PMA distribution at high irradiation dose, have important implications for the applications of PMA materials in data storage and spin-electronic applications which require tight control of the anisotropy, anisotropy distributions and resonant behavior.  

\section{Acknowledgments}
We thank Gabriel Chaves for help in fitting the data to the TMS model. This work was supported by NSF Grant No. DMR-0706322.
\bibliography{others}
\end{document}